# Structural instabilities of infinite-layer nickelates from *first-principles* simulations

Álvaro Adrián Carrasco Álvarez[1,2], Sébastien Petit[1], Lucia Iglesias[2], Wilfrid Prellier[1], Manuel Bibes[2] and Julien Varignon[1]

[1] CRISMAT, Normandie Université, ENSICAEN, UNICAEN, CNRS, 14000 Caen, FRANCE
[2] Unité Mixte de Physique, CNRS, Thales, Université Paris Saclay, 91767 Palaiseau, France

## Abstract

Rare-earth nickelates $RNiO_2$ adopting an infinite-layer phase show superconductivity once La, Pr or Nd are substituted by a divalent cation. Either in the pristine or doped form, these materials are reported to adopt a high symmetry, perfectly symmetric, $P_4/mmm$ tetragonal cell. Nevertheless, bulk compounds are scarce, hindering a full understanding of the role of chemical pressure or strain on lattice distortions that in turn could alter magnetic and electronic properties of the 2D nickelates. Here, by performing a full analysis of the prototypical $YNiO_2$ compound with first-principles simulations, we identify that these materials are prone to exhibit $O_4$ group rotations whose type and amplitude are governed by the usual R-to-Ni cation size mismatch. We further show that these rotations can be easily tuned by external stimuli modifying lattice parameters such as pressure or strain. Finally, we reveal that H intercalation is favored for any infinite-layer nickelate member and pushes the propensity of the compounds to exhibit octahedra rotations.

The nickelate oxides have long been proposed as potential candidates for high temperature superconductivity [1,2]. However, superconductivity was only achieved recently after the RNiO$_3$ phase was doped with a divalent cation (Ca or Sr) and transformed into a RNiO$_2$ infinite-layer structure (R=La, Pr, Nd) by reduction with CaH$_2$ [3–9]. Although the origin of superconductivity is yet to be clarified in these compounds, it was originally pointed that *4f* electrons in Nd$_{0.8}$Sr$_{0.2}$NiO$_2$ may play a key role [10]. This is however ruled out by the appearance of superconductivity in doped LaNiO$_2$ in which there are no *4f* electrons. Thus, the occurrence of superconductivity in different infinite-layer nickelates suggests that this property may appear in any member of the family independently of the rare-earth and of the presence of *4f* electrons. So far, all observed superconducting nickelates (*i.e.* R=La, Pr or Nd), exclusively obtained as thin films, are reported to crystallize within a highly symmetric, undistorted, *P$_4$/mmm* cell consisting of NiO$_2$ layers intercalated between rare-earth layers [11–14] (see Fig. 1.a). Nevertheless, there are no experimental reports of infinite-layer nickelates with other R cations either in the bulk or thin film form to the best of our knowledge. One may thus wonder whether the *P$_4$/mmm* structure is persistent across the whole family or if A-to-B cation size mismatch appearing in ABO$_3$ perovskites – quantified through the Goldschmidt tolerance factor [15] and yielding the usual octahedra rotations and potential gap opening in such compounds [16]— may appear in infinite-layer nickelates with small R cations.

Recent references have addressed this question from a theoretical point of view using the phonon dispersion curve of the infinite layer nickelates [17,18]. The authors identified the existence of a single O$_4$ group rotation yielding a a$^0$a$^0$c$^-$ rotation pattern ($\emptyset_z^-$ mode in Fig. 1.c) following Glazer's notation [19] in nickelate members with a small R cation such as YNiO$_2$. Furthermore, YNiO$_2$ was also proposed to disfavor H contamination that can potentially appear during the CaH$_2$ reduction process of the parent RNiO$_3$ phase [20,21]. Nevertheless, that study was guided by phonon calculations and neglects the potential lattice mode couplings that can help stabilizing different tilt patterns and/or antipolar motion of octahedra cations [16,22–24] such as those exhibited by the usual *Pbnm* symmetry of ABO$_3$ perovskites ($\emptyset_z^+$, $\emptyset_{xy}^-$ and $A_p$ modes displayed in Figure 1).

We performed *first-principles* simulations using Density Functional Theory (DFT) aiming at identifying the ground state structure of various nickelate members. We identify that compounds with small R cations such as YNiO$_2$ are prone to exhibit O$_4$ rotations yielding an orthorhombic *Pbnm* symmetry of the form $a^-a^-c^+$. This behavior is reminiscent of the octahedra tilt pattern exhibited by most $ABO_3$ perovskite, including the parent $RNiO_3$ phase – at the exception of $LaNiO_3$ [25]. In these members, a trilinear term between the $a^-a^-c^0$ ($\phi_{xy}^-$) and $a^0a^0c^+$ ($\phi_z^+$) rotations and an antipolar motion of R cations, allowed in the free energy expansion, lowers the total energy and forces their appearance in the ground state. Upon increasing the rare-earth radius such as in $GdNiO_2$, the $O_4$ groups tilt patterns change to $a^0a^0c^-$ within a *I4/mcm* symmetry. This is induced by an unstable phonon mode overcoming the energy gain associated with the alternative tilt pattern yielding the *Pbnm* symmetry. Structural distortions are absent for $PrNiO_2$ and $LaNiO_2$ and these adopt the high symmetry, totally undistorted, *P4/mmm* cell. Finally, starting from the proper ground state structure, all materials prefer to stabilize hydrogen in the unit cell that in turn favor octahedra rotations and open a finite band gap.

***First-principles* Density Functional Theory** (DFT) simulations are performed with the *Vienna Ab initio Simulation Package* (VASP) [26,27] using the recent Strongly Correlated and Appropriately Normalized (SCAN) functional [28] in order to better cancel self-interaction errors inherent to practiced DFT. It was previously shown to be well-suited for $ABO_3$ compounds with a *3d* element in bulk [29], including to treat doping effects in rare-earth nickelates [30] or in cuprates [31,32]. The explored structures are based on different tilt patterns of $O_4$ motifs that are defined using the Glazer's notation [19]: it entails $a^0a^0c^+$ (*P4/mbm*, $\phi_z^+$ mode), $a^-a^-c^0$ (*Imma*, $\phi_{xy}^-$ mode), $a^0a^0c^-$ (*I4/mcm*, $\phi_z^-$ mode), $a^-a^-c^-$ (*C2/c*, $\phi_z^- + \phi_{xy}^-$ modes) and $a^-a^-c^+$ (*Pbnm*, $\phi_z^+ + \phi_{xy}^-$ modes) tilt patterns. All calculations are performed using a ($\sqrt{2}, \sqrt{2}, 2$) supercell (4 formula units) with respect to the primitive, high symmetry *P4/mmm* cell. $RNiO_2$ tested compounds cover R=Y, Gd, Pr and La. Structural relaxation is performed until forces acting on each atom are lower than 1 meV/Å. The energy cut-off is set to 650 eV and the k-mesh is set to 8x8x6 points for the 4 f.u cell. *4f* electrons are treated in the simulations and the magnetic order on both A and Ni cations is restricted to the AFM-C order previously identified with the SCAN functional [33].

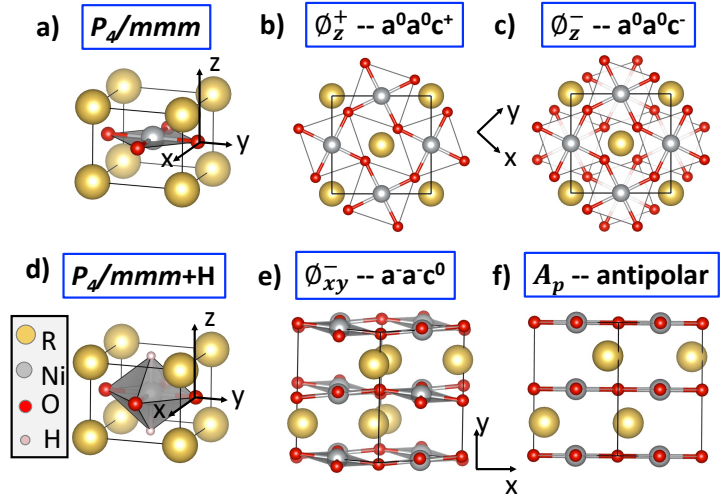

*Figure 1: Sketch of the different lattice distortions that can appear in the 2D nickelates. a) Highly symmetric undistorted P4/mmm cell adopted by nickelates. b) and c) In-phase $a^0a^0c^+$ (a) and anti-phase $a^0a^0c^-$ (b) $O_4$ group rotations. d) H intercalation at the apex of $O_4$ groups. e) Antiphase $a^-a^-c^0$ rotation. f) Antipolar motion $A_p$ of R cations. We note that the antiphase rotation $a^-a^-c^0$ is also accompanied by an antipolar motion of A site cations.*

**The structural relaxation in $YNiO_2$:** We first focus on $YNiO_2$ for which the parent perovskite phase $YNiO_3$ would show the largest A-to-B cation size mismatch among our considered compounds. Thus, one expects a larger propensity to exhibit $O_4$ group rotations. We summarize in Table 1 the total energy difference between all tested tilt patterns with respect to the undistorted high symmetry $P_4/mmm$ cell. The *Imma* ($a^-a^-c^0$) and $P_4/mbm$ ($a^0a^0c^+$) cells are found unstable and relax back to a $P_4/mmm$ cell. It suggests that these tilt patterns alone are not stable in $YNiO_2$. For the other possibilities, we observe that $YNiO_2$ is nevertheless clearly prone to exhibit oxygen group rotations since the undistorted primitive $P_4/mmm$ cell is less stable than all other tested cells. Among all tilt patterns, we identify the orthorhombic *Pbnm* symmetry showing the usual $a^-a^-c^+$ $O_4$ group tilt patterns as the structural ground state. Our identification of an orthorhombic symmetry is in contrast with two recent DFT studies that found the $I_4/mcm$ cell to be the structural ground state [17,18]. This discrepancy could be originated from a lower level of approximation of exchange-correlation (xc) phenomena through the used DFT xc functionals (*i.e.* PBEsol [34] or PBE [35] functionals). Nevertheless, we have tested the relative stability between *Pbnm* and $I_4/mcm$ symmetries with PBEsol and PBEsol+U (U=5 eV on Ni *3d* orbitals [36]) and still find that the *Pbnm* cell is lower in energy than the $I_4/mcm$ (ΔE=-27 and -57 meV/f.u. for the energy difference, respectively).

| symmetry | P$_4$/mmm | I$_4$/mcm | C2/c | Pbnm |
|---|---|---|---|---|
| ΔE (meV/f.u) | 0 | -175 | -179 | -212 |
| $\emptyset_z^+$ (a$^0$a$^0$c$^+$ - M$_2^+$) | - | - | - | 0.593 |
| $\emptyset_{xy}^-$ (a$^-$a$^-$c$^0$ - A$_5^-$) | - | - | 0.156 | 0.333 |
| $\emptyset_z^-$ (a$^0$a$^0$c$^-$ - A$_4^-$) | - | 0.504 | 0.517 | - |
| A$_p$ (Z$_5^-$) | - | - | - | 0.449 |

**Table 1**: *Total energies differences (in meV/f.u) with respect to the P$_4$/mmm cell for the different tilt patterns that could be stabilized in YNiO$_2$. Amplitudes of structural distortions (in Å/f.u.) of the different relaxed structure with respect to a high symmetry P$_4$/mmm cell. Amplitudes are extracted from a symmetry mode analysis using Amplimodes software [37,38].*

***An avalanche effect produces the Pbnm cell in YNiO$_2$:*** Aiming at understanding the origin of the orthorhombic *Pbnm* cell exhibited by YNiO$_2$, we first perform a symmetry mode analysis of our stabilized structures with respect to the high symmetry *P$_4$/mmm* unit cell using the Amplimode software [37,38] (see Table 1). We see that the *I$_4$/mcm* and *C$_2$/c* cells only exhibit anti-phase rotations whereas the *Pbnm* symmetry shows the a$^0$a$^0$c$^+$ ($\emptyset_z^+$) and a$^-$a$^-$c$^0$ ($\emptyset_{xy}^-$) plus the antipolar motion *A$_p$* of A site cations (Fig 1.f) identified in any *Pbnm* cell of ABO$_3$ materials [16,23,24,39]. We then plot the potential energy surface associated with individual distortion modes, starting from the high symmetry cell *P$_4$/mmm* cell with lattice parameters of the relaxed *P$_4$/mmm* cell (Fig. 2 a). Amazingly, as one can see, only the a$^0$a$^0$c$^-$ ($\emptyset_z^-$) mode presents a double well potential while all other modes such as a$^-$a$^-$c$^0$ ($\emptyset_{xy}^-$) and a$^0$a$^0$c$^+$ ($\emptyset_z^+$) exhibit a single well potential. In other words, only the $\emptyset_z^-$ mode is unstable and is willing to appear spontaneously in the material. This result is in agreement with the results of Bernardini *et al* [17] as well as Xia *et al* [18] that found a single instability associated with the a$^0$a$^0$c$^-$ rotation in the phonon dispersion curve of YNiO$_2$.

So the next question is: "*how can the a$^0$a$^0$c$^+$ and a$^-$a$^-$c$^0$ tilt pattern be more stable than the a$^0$a$^0$c$^-$ tilt pattern in this compound if these two former modes are not unstable by themselves*"? In order to understand the stabilization of the *Pbnm* cell, we develop the free energy $F$ associated with the distortions appearing the ground state structure (*i.e.* $\emptyset_{xy}^-$, $\emptyset_z^+$, A$_p$) around the *P$_4$/mmm* cell:

$$F(Pbnm) \propto \alpha(\emptyset_z^+)^2 + \beta(\emptyset_z^+)^4 + \gamma(\emptyset_{xy}^-)^2 + \delta(\emptyset_{xy}^-)^4 + \lambda(A_p)^2 + \varsigma(A_p)^4 + \xi\emptyset_z^+\emptyset_{xy}^-A_p \quad (eq.1)$$

While even terms are expected in the free energy expansion of eq.1, we surprisingly recover the trilinear term (*i.e.* an odd term) between $\emptyset_z^+$, $\emptyset_{xy}^-$ and $A_p$ modes identified in ABO$_3$ perovskites adopting a *Pbnm* cell [16,22–24,40]. This term is only allowed within the *Pbnm* symmetry – or lower symmetries – and is absent in structures adopting a$^0$a$^0$c$^-$ or a$^-$a$^-$c$^-$ rotation patterns (*i.e. I$_4$/mcm* or *C$_2$/c* cells). The trilinear term has a surprising role in infinite-layer nickelates: by plotting the potential energy surface as a function of A$_p$ but at fixed amplitudes of $\emptyset_z^+$ and $\emptyset_{xy}^-$ distortions, we see that upon increasing amplitudes of $\emptyset_z^+$ and $\emptyset_{xy}^-$ distortions, the energy minimum is shifted to lower energies and finite amplitudes of the A$_p$ mode (see Figure 2.d). Although the $\emptyset_z^+$, $\emptyset_{xy}^-$ and $A_p$ modes are all stable individually with respect to the *P$_4$/mmm* cell, the trilinear term possesses a strong and negative contribution to *F* in eq.1 that can produce the concomitant appearance of all the three distortions despite these being not unstable by themselves. Thus, the *Pbnm* cell exhibited by YNiO$_2$ appears through an avalanche effect of *a priori* stable modes. This lattice mode coupling then explains the discrepancy between our results and those of Refs. [17,18], in which lattice mode couplings were ignored.

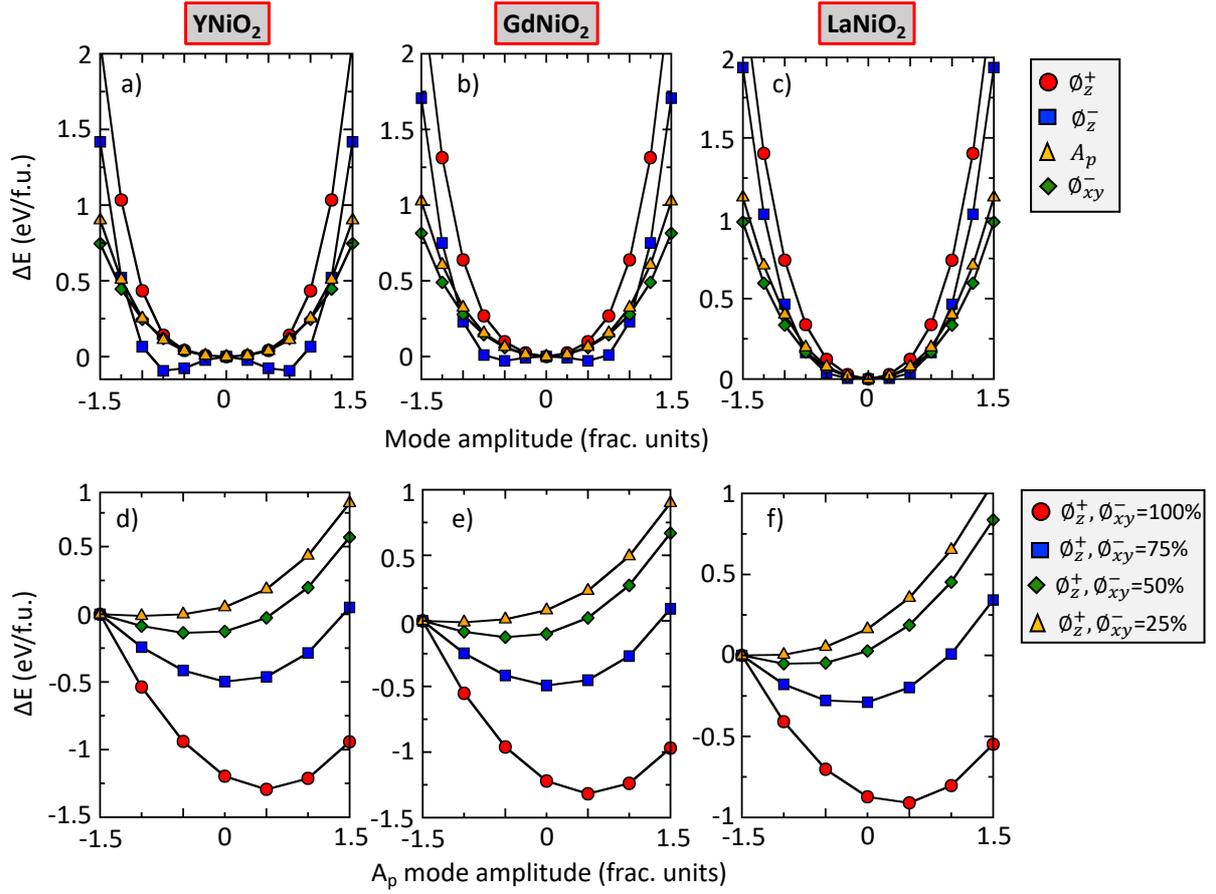

**Figure 2:** Energy difference (in meV/f.u.) with respect to the $P_4/mmm$ cell associated with the different lattice distortions that can appear in the 2D nickelates tested in the simulations. a-c) Potential energy surface associated for modes alone in $YNiO_2$, $GdNiO_2$ and $LaNiO_2$. d-f) Potential energy surface associated with the antipolar mode $A_p$ at fixed amplitude of the octahedra rotations.

**Lattice relaxation effect on the stability of the Pbnm cell in $YNiO_2$**: Although the identified trilinear term does explain the concomitant appearance of the $a^-a^-c^+$ $O_4$ groups rotations by producing a large energy gain, it does not imply that it is the global minimum of the energy landscape as a function of the lattice distortion. Indeed, individual modes produce a large and positive contribution to the total energy that can counterbalance the negative contribution from the trilinear term of eq.1 as inferred by Figure 2.a. Using the optimized $P_4/mmm$ lattice parameter, we indeed observe that the $Pbnm$ cell is not the ground state, this phase being higher in energy than the $I_4/mcm$ cell by $\Delta E=+18.5$ meV/f.u. Allowing lattice parameters to relax in both symmetries, the cell can distort and adjust the rotation pattern. Consequently, a substantial energy gain is achieved by minimizing strain effects in the $Pbnm$ symmetry thereby allowing it to become 37 meV/f.u. lower than the $I_4/mcm$ phase. One notices that if we start from a $P_4/mmm$ cell but at the $Pbnm$ cell volume, the potential energy surface of the $\emptyset_z^+$ mode shows a double well potential thereby further lowering the total energy of the compound.

***Epitaxial strain effect in YNiO₂***: Most of the infinite-layer nickelates realized so far are stabilized in thin film form. Having established that cell distortions are a key parameter to reach the *Pbnm* cell in YNiO₂, one may thus wonder whether epitaxial strain effects may alter the stability of the *Pbnm* cell with respect to the *I₄/mcm* cell. Imposing the in-plane lattice constant of a relaxed cubic SrTiO₃ with the SCAN functional ($a_{STO}$=3.908 Å), we find that the *Pbnm* cell remains the ground state although its stability with respect to the *I₄/mcm* symmetry strongly diminishes (ΔE=-10 meV/f.u.). We conclude here that strain is likely to disfavor the orthorhombic *Pbnm* cell.

***Chemical pressure effect on the structural phase transition of other nickelate members***: Since octahedra rotations amplitude are governed by steric effects in ABO₃ perovskites, one may also wonder if these phenomena do occur in the infinite-layer nickelates. We performed structural relaxation for a few other compounds such as GdNiO₂, PrNiO₂ and LaNiO₂, *i.e.* compounds with increasing A site cation radius. We identify that GdNiO₂ adopts a *I₄/mcm* ground structure associated with only the $a^0a^0c^-$ O₄ group rotation (ΔE=-30 meV/f.u. with respect to the *Pbnm* cell) while PrNiO₂ and LaNiO₂ exhibit a purely undistorted *P₄/mmm* cell as a ground state. In all these compounds, the trilinear term linking the in-phase $\emptyset_z^+$ and antiphase $\emptyset_{xy}^-$ rotation with an antipolar motion $A_p$ remains effective as shown in Figure 2.e and in fact this trilinear term in eq.1 may always produce a negative contribution to the total energy whatever the material. However, its energy gain cannot overcome the large positive contribution to the total energy of the individual $\emptyset_z^+$, $\emptyset_{xy}^-$ and $A_p$ modes that still exhibit single wells (Figs. 2.b and c). As a result, GdNiO₂ only exhibits an antiphase octahedra rotation around the c axis ($a^0a^0c^-$) that arises from an unstable in the *P₄/mmm* cell as inferred from Fig 2.b. In LaNiO₂, all distortion modes are found to be stable, including the antiphase $a^0a^0c^-$ rotation (Figure 2.c). We conclude here that, akin to ABO₃ compounds, A-to-B cation size mismatch is still effective in reducing the crystal symmetry in the infinite-layer phase.

***H intercalation is found to be favored in all compounds:*** Since, experimentally, infinite-layer nickelates are obtained by a CaH₂ chemical reduction of the parent RNiO₃ perovskite phase, it is expected that some level of H intercalation may be present. Thus, we now examine the influence of this effect on phase stability. In NdNiO$_2$, it was proposed that the pristine phase is highly unstable while its hydrogenation makes it more stable [21]. Recently, it was discussed that going to YNiO$_2$ [17] or decreasing the (ab)-plane $P_4/mmm$ lattice parameter could disfavor H intercalation inside these nickelates [20]. These possibilities thereby offer a knob to get rid of H intercalation in the materials that ultimately can alter their properties. We have computed the binding energy for H intercalation at the apex of O$_4$ groups, which was found to be the preferred position in Refs. [17,21] (see Fig. 1.d), considering the extreme limit of one H intercalated per primitive unit cell (*i.e.* RNiO$_2$H) using the following equation:

$$E_{binding} = E(RNiO_2H) - E(RNiO_2) - E(H) \quad (eq.2)$$

In eq.2, we consider total energies obtained after structural relaxation for RNiO$_2$H and RNiO$_2$. The reference for the total energy of hydrogen is set to half the total energy of relaxed H$_2$ with the SCAN functional. From eq.2, it follows that a negative (positive) value for $E_{binding}$ indicates a favorable (unfavorable) Hydrogen intercalation in the material.

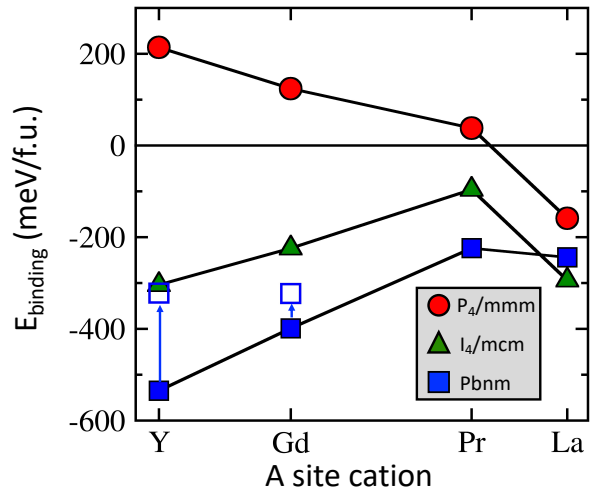

*Figure 3: Binding energy (in meV/f.u.) for hydrogen intercalation in the different 2D nickelates. The reference energy is set to pristine material with the $P_4/mmm$ space group. Arrows correspond to the binding energy for the hydrogenated Pbnm cell computed with respect to the ground state structure of the pristine phase for YNiO$_2$ and GdNiO$_2$ (blue open squares).*

Trends for $E_{binding}$ as a function of the rare-earth are reported in Figure 3. Using the pristine $P_4/mmm$ structure as a reference energy for all nickelate members, the binding energy for intercalating hydrogen is positive for $YNiO_2$ and then decreases upon increasing the radius of the A site cation, until it becomes negative only for $LaNiO_2$. This is compatible with the result of Ref. [20] that found H intercalation to be favorable in $LaNiO_2$. Thus, from the strict point of view of the high symmetry, undistorted $P_4/mmm$ cell, $LaNiO_2$ would be the only member to favor stabilization of H inside the cell during the $CaH_2$ chemical reduction process. Nevertheless, we have tested the $I_4/mcm$ and $Pbnm$ symmetries with H intercalated and identified that both distorted cells produce large energy gains with respect to the $P_4/mmm$ cell that in turn strongly favors H intercalation. The result remains valid even if one considers the pristine $Pbnm$ or $I_4/mcm$ ground state as a reference energy for $YNiO_2$ and $GdNiO_2$, respectively (see rescaled values indicated by arrows in Fig. 3). Consequently, unlike results from Ref. [17] where H inclusion is proposed to be ruled out in $YNiO_2$ – shall it adopt the $P_4/mmm$ symmetry –, we clearly identify that once using the proper distorted cell for each nickelate member, all nickelates have negative binding energies towards the insertion of H, with a mean value of -290meV/f.u across the series. Although our simulations are restricted to the extreme limit of 1 H inserted per primitive cell, it is therefore very likely that fractions of H tend to be trapped inside the infinite layer nickelate whatever the A site cation.

Several side effects of the inclusion of hydrogen in the material have to be pointed out: (i) intercalating H in all nickelate members favor the appearance of octahedra rotation with notably a tilt pattern around the c axis that is reminiscent of the parent $RNiO_3$ phase with a $a^0a^0c^+$ rotations for all members except for R=La in which the $a^0a^0c^-$ rotation is found more stable; (ii) H acts as an acceptor [21] and (iii) favors the appearance of a high spin $Ni^{2+}$ state with a magnetic moment of 1.5 $\mu_B$ – it is lower than the expected value of 2 $\mu_B$ due to spillage on surrounding O and H atoms. All these effects have the tendency to open a small, but finite, band gap in all the nickelates that are otherwise all found metallic in the undoped phase [21].

***Conclusions***: We have identified that steric effects play an important role in the structure of infinite-layer nickelates and that most members will show $O_4$ groups rotation producing either the usual orthorhombic $Pbnm$ symmetry for A site cations with small radius or an $I_4/mcm$ symmetry for A site cations with moderate radius. Only members with the largest rare-earth (R=La, Pr, Nd), for which superconductivity is reported so far, adopt a perfectly undistorted cell in the pristine phase. Finally, we have clarified that these rotations favor H intercalation in the material that in turn increases rotations amplitude, ultimately resulting in an insulating phase with only $Ni^{2+}$ cations in a high spin state. This hints at the fact that a fraction of H may be intercalated and that doping with a divalent cation may be required to annihilate the incipient insulating behavior as observed in the resistivity curve of undoped $NdNiO_2$ compound [3].

*Acknowledgements:* This work has received financial support from the CNRS through the MITI interdisciplinary programs under the project SuNi. Authors acknowledge access granted to HPC resources of Criann through the projects 2020005 and 2007013 and of Cines through the DARI project A0080911453. L. I. acknowledges the funding from the Ile de France region and the European Union's Horizon 2020 research and innovation programme under the Marie Sklodowska-Curie grant agreement №21004513 (DOPNICKS project).